\documentclass[submission,copyright,creativecommons]{eptcs}
\providecommand{\event}{ICLP 2026} 

\usepackage{iftex}
\usepackage{listings}
\usepackage{graphicx}
\usepackage{framed}
\usepackage{xcolor}
\usepackage{subcaption}
\usepackage{booktabs}
\usepackage{asplisting}
\usepackage{multirow}

\newcommand{\ucap}{\ensuremath{\mathit{UCAP}}}
\newcommand{\iucap}{\ensuremath{\mathit{IUCAP}}}

\ifpdf
  \usepackage{underscore}         
  \usepackage[T1]{fontenc}        
\else
  \usepackage{breakurl}           
\fi

\title{Streamliners for Answer Set Programming}
\author{Florentina Voboril
\institute{TU Wien\\Vienna, Austria}
\email{florentina.voboril@tuwien.ac.at}
\and
Martin Gebser
\institute{University of Klagenfurt\\Klagenfurt, Austria}
\email{Martin.Gebser@aau.at}
\and
Stefan Szeider
\institute{TU Wien\\Vienna, Austria}
\email{sz@ac.tuwien.ac.at}
\and
Alice Tarzariol
\institute{University of Klagenfurt\\Klagenfurt, Austria}
\email{Alice.Tarzariol@aau.at}
}
\def\titlerunning{Streamliners for Answer Set Programming}
\def\authorrunning{F. Voboril, M. Gebser, S. Szeider \& A. Tarzariol}
\begin{document}
\maketitle

\begin{abstract}
Streamliner constraints reduce the search space of combinatorial problems by
ruling out portions of the solution space. We adapt the StreamLLM approach,
which uses Large Language Models (LLMs) to generate streamliners for Constraint
Programming, to Answer Set Programming (ASP). Given an ASP encoding and a few
small training instances, we prompt multiple LLMs to propose candidate
constraints. Candidates that cause syntax errors, render satisfiable instances
unsatisfiable, or degrade performance on all training instances are discarded.
The surviving streamliners are evaluated together with the original encoding, and we report results for a virtual best encoding (VBE) that, for each instance, selects the fastest among the original encoding and its streamlined variants.
On three ASP competition benchmarks (Partner Units Problem, Sokoban, Towers of
Hanoi), the VBE achieves speedups of up to 4--5$\times$ over the original
encoding. 
\end{abstract}

\section{Introduction}

Answer Set Programming (ASP) has emerged 
as a powerful paradigm for
declarative problem solving, with successful applications in planning, scheduling,
configuration, and knowledge representation~\cite{CalimeriFGIKKLM20}.
Modern ASP solvers like \textsc{clingo}~\cite{DBLP:journals/tplp/GebserKKS19}
combine grounding with conflict-driven clause learning, enabling
the solution of hard combinatorial problems. However, solver performance
remains highly sensitive to problem encodings: two logically
equivalent formulations can lead to vastly different grounding sizes and solving
times.

A well-known technique for improving solver performance is the addition of
\emph{streamliner constraints} \cite{GomesSellmann2004}. These are auxiliary
constraints intended to reduce the search space by ruling out parts of the
solution space. They need not be logically implied by the original encoding and
may remove some solutions, but should preserve satisfiability for the instances
of interest. Streamliners include \emph{symmetry-breaking
constraints}~\cite{Walsh2012}, which remove equivalent solutions arising from
problem symmetries, and \emph{implied constraints}, which make implicit problem
structure explicit to the solver. In the Constraint Programming (CP) community, Spracklen et al.~\cite{SpracklenDAM2023}
recently demonstrated that portfolios of automatically generated streamliners
can match or exceed hand-crafted constraints. In ASP, systems like
\textsc{sbass}~\cite{drtiwa11a} and \textsc{BreakID}~\cite{debo2016a}
automatically generate propositional symmetry-breaking constraints, while
inductive logic programming methods such as
conflict-driven constraint learning~\cite{Law2023} and core-guided
reformulation~\cite{LeoGBW2024} derive first-order constraints from small training
instances~\cite{tagesc22}.
These methods either operate at the propositional level (often
causing grounding bottlenecks on larger instances) or require carefully designed
language biases (limiting generality).

Voboril et al.~\cite{voboril-jair-2025} recently demonstrated that Large Language Models (LLMs)
can generate streamliner constraints for CP
models. Their \emph{StreamLLM} approach prompts an LLM with a problem encoding
and asks it to propose constraints that reduce the search space. Candidate
constraints are validated on small training instances and the best-performing
ones are retained. This approach requires no hand-crafted language bias or
domain templates, but only uses the encoding plus a few small training instances.%

The 
combination of LLMs and ASP has been explored along two directions.
Neural-symbolic hybrids like NeurASP~\cite{YangIL2020} and differentiable ASP
solvers~\cite{SkryaginSL2024} integrate neural components with ASP reasoning,
but assume that the logic program is manually written. For automated program
generation, Ishay et al.~\cite{IshayYL2023} prompt LLMs to produce complete ASP programs
from natural language (NL), and Coppolillo et al.~\cite{CoppolilloEtAl2024} fine-tune LLMs for ASP
syntax. However, one-shot generation often yields flawed
programs~\cite{SantanaBCF24}. Agentic approaches address this through iterative
refinement with solver feedback: the MCP-Solver~\cite{Szeider2025mcp} exposes
ASP capabilities to LLMs via the Model Context Protocol, and evaluation on
ASP-Bench~\cite{SzeiderASPBench2025} shows that feedback-driven agents 
succeed on benchmark problems. Our work differs from NL-to-ASP translation as we do
not generate entire programs, but propose auxiliary constraints that augment
existing encodings.

In this paper, we investigate whether LLM-based streamliner generation can be
adapted to ASP, which presents distinct challenges
than CP: the non-monotonic semantics of stable models means that adding
constraints or auxiliary definitions can have subtle effects on solution sets.
LLMs may generate not only constraints but also helper
rules that introduce new atoms, which can interact non-trivially with the
minimality condition of stable models. The rich syntax of ASP
(aggregates, choice rules, conditional literals, optimization statements) also requires LLMs to generate
syntactically valid and semantically meaningful code.
We adapt the StreamLLM pipeline to ASP by
$(1)$ prompting multiple frontier LLMs to generate candidate streamliners for a given  ASP encoding,
$(2)$ filtering candidates that cause syntax errors, make satisfiable
          training instances unsatisfiable, or degrade runtime performance
          on all training instances,
$(3)$ selecting the best-performing streamliners based on runtime
          improvements, and 
$(4)$ evaluating the original encoding together with the streamlined variants 
as well as the virtual best encoding (VBE) for each instance,
which serves as an oracle-style analysis of complementarity.

We evaluate our approach on three benchmark problems from the ASP competition
2011~\cite{calimeri2011third} - Partner Units Problem, Sokoban, and
Towers of Hanoi - observing speedups of up to 4--5$\times$ over the original encoding.
The observation that 
different LLMs and different runs produce semantically diverse constraints,
not simple syntactic variations, 
suggests that 
our approach captures genuine problem structure rather than exploiting
superficial patterns.

The rest of this paper is structured as follows.
Section~\ref{sec:preliminaries} provides background on ASP and related work.
Section~\ref{sec:method} describes our adaptation of StreamLLM to ASP.
Section~\ref{sec:experiments} presents the experiments and analyzes the generated streamliners.
Section~\ref{sec:conclusion} concludes 
by discussing limitations and future directions.

\section{Preliminaries}
\label{sec:preliminaries}

This section provides brief overviews of ASP and previous works related to streamliner constraints.

\subsection{Answer Set Programming}

We consider first-order ASP programs in the modeling language of \textsc{clingo}
\cite{CalimeriFGIKKLM20,gekakasc12a}. That is, \textit{terms} include
(i) \textit{constants}, which can be tokens starting with a lowercase letter, integers, strings enclosed in double quotes, 
the least term \lstinline{#inf} or the greatest term \lstinline{#sup},
(ii) \textit{variables}, i.e., tokens starting with an uppercase letter or the anonymous variable ``\lstinline{_}'', and
(iii) \textit{functions} of the form
\lstinline{f(t}$_1$\lstinline{,}$\dots$\lstinline{,t}$_n$\lstinline{)}, for a function symbol \lstinline{f} (a token starting with a lowercase letter) and terms
\lstinline{t}$_1,\dots,{}$\lstinline{t}$_n$,
tuples \lstinline{(t}$_1$\lstinline{,}$\dots$\lstinline{,t}$_n$\lstinline{)},
or arithmetic expressions composed of integers, variables, operators
``\lstinline{+}''${},{}$``\lstinline{-}''${},{}$``\lstinline{*}''${},{}$``\lstinline{/}''${},\mbox{ etc.}$ and parentheses.

\textit{Atoms} can be of three forms:
(i) \textit{symbolic} atoms
\lstinline{p(t}$_1$\lstinline{,}$\dots$\lstinline{,t}$_n$\lstinline{)},
for a predicate symbol \lstinline{p} (a token starting with a lowercase letter) and terms
\lstinline{t}$_1,\dots,$\lstinline{t}$_n$, which are
just written \lstinline{p} in case $n=0$,
(ii) \textit{built-in} atoms
\lstinline{t}$_1 \circ {}$\lstinline{t}$_2$ composed of terms
\lstinline{t}$_1,{}$\lstinline{t}$_2$ and a comparison operator
${\circ}\in \{
 $\lstinline{<}${},{}$%
  \lstinline{<=}${},{}$%
  \lstinline{>=}${},{}$%
  \lstinline{>}${},{}$%
  \lstinline{=}${},{}$%
  \lstinline{!=}$\}$, and
(iii) \textit{aggregate} atoms
\lstinline{t}$_1 \circ_1 {}$\lstinline{#}$\mathit{aggr}$%
\lstinline|{|$e_1;\dots;e_k$\lstinline|}|${}\circ_2{}$\lstinline{t}$_2$,
where \lstinline{t}$_1,{}$\lstinline{t}$_2$ are terms,
${\circ}_1,{\circ}_2$ are comparison operators,
\lstinline{#}$\mathit{aggr}\in\{
 $\lstinline{#count}${},{}$%
  \lstinline{#sum}${},{}$%
  \lstinline{#min}${},{}$%
  \lstinline{#max}$\}$ is an \textit{aggregate function},
and each $e_i$, for $1\leq i\leq k$,
is an \textit{aggregate element} 
\lstinline{t}$_{1_i}$\lstinline{,}$\dots$\lstinline{,t}$_{n_i}$%
\lstinline{:} $\ell_{1_i}$\lstinline{,}$\dots$\lstinline{,}$\ell_{m_i}$
over terms
\lstinline{t}$_{1_i},\dots,{}$\lstinline{t}$_{n_i}$ and \textit{basic literals}
$\ell_{1_i},\dots,\ell_{m_i}$ such that each $\ell_{j_i}$, for $1_i\leq j_i\leq m_i$, is
$a$, \lstinline{not} $a$ or \lstinline{not} \lstinline{not} $a$ for some symbolic or built-in atom $a$.
Basic literals as well as $a$, \lstinline{not} $a$ and \lstinline{not} \lstinline{not} $a$ over an aggregate atom $a$ are \emph{literals}.
A \textit{choice element} $c_i$ is an aggregate element
\lstinline{t}$_{1_i}$
\lstinline{:} $\ell_{1_i}$\lstinline{,}$\dots$\lstinline{,}$\ell_{m_i}$
such that the (constant or function) term \lstinline{t}$_{1_i}$ has the form of a symbolic atom, and 
\lstinline{t}$_1 \circ_1 {}$%
\lstinline|{|$c_1;\dots;c_k$\lstinline|}|${}\circ_2{}$\lstinline{t}$_2$
with choice elements $c_1,\dots,c_k$ is a \textit{choice}.
Either or both of the comparison operators ${\circ}_1,{\circ}_2$ can be omitted for a choice or an aggregate atom, in which case they default to ``\lstinline{<=}'',
and if \lstinline{t}$_1$ or \lstinline{t}$_2$ is skipped in addition,
\lstinline{#inf} resp.\ \lstinline{#sup} is taken as the corresponding default value.

An ASP \textit{program} is a set of \textit{rules} of the form
$h$ \lstinline{:-} $b_1$\lstinline{,}$\dots$\lstinline{,}$b_m$\lstinline{.},
where $h$ is a symbolic atom or a choice, and
$b_1,\dots,b_m$ are literals.
The rule head $h$ may be omitted, in which case we call the rule a \textit{constraint}.
In order to be processed by \textsc{clingo}, an ASP program has to be \emph{safe} \cite{CalimeriFGIKKLM20}, roughly meaning that all variables must have positive occurrences (outside of arithmetic expressions) in symbolic atoms belonging to the rule body $b_1$\lstinline{,}$\dots$\lstinline{,}$b_m$\lstinline{.}
The semantics of an ASP program is given by \textit{stable models} \cite{gehakalisc15a}, which are sets of (true) ground atoms that satisfy and are derivable from the instances of the program's (first-order) rules.
We say that an ASP program is \textit{satisfiable} if it has some stable model, and \textit{unsatisfiable} otherwise.
Proper first-order rules are called an ASP \textit{encoding}, and the rules without variables a problem \textit{instance}.

\subsection{Related Works}

Although the term streamliner is not widely used in the ASP literature, many
works have investigated the identification of constraints that prune the search
space to improve solving performance. Regarding symmetry breaking, research in
ASP has focused on automatically producing \textit{symmetry-breaking
constraints} (SBCs), inspired by techniques from the Boolean satisfiability (SAT) community that generate
propositional SBCs based on properties of permutation
groups~\cite{sakallah09a}. The systems \textsc{sbass}~\cite{drtiwa11a} and
\textsc{BreakID}~\cite{debo2016a} analyse a ground ASP program to identify
symmetries and add propositional SBCs, which can speed up solving on small and highly 
combinatorial instances, but often does not scale to large ones.
The work
of Devriendt et al.~\cite{debobrde16a} analyses predicates for interchangeable argument values
and adds propositional SBCs imposing lexicographical orderings; however, this
approach handles only local domain symmetries. Moreover, propositional SBCs can
negatively affect solving performance through redundant constraints, and users
gain no insight into the symmetry structure.

These shortcomings are addressed by Tarzariol et al.~\cite{tagesc22}, who exploit symmetries
identified by \textsc{sbass} to define an inductive logic programming task that
learns first-order constraints 
for both decision problems~\cite{tagesc22b} and optimization
problems~\cite{tagesc23}. The constraints are obtained by lifting propositional
symmetries from small representative instances and are validated on a test set, but
without formally proving them; thus, this approach generates streamliners that are not
necessarily SBCs. 
The use of inductive logic programming requires providing a language bias specifying
building blocks for the constraints, which makes learning efficient constraints
almost as challenging as optimizing an encoding by hand.

Domain-specific streamliners have been developed for particular applications:
Yli-Jyr{\"{a}} et al.~\cite{yli2023pruning} incorporate pruning constraints into preventive
maintenance scheduling, while Hus{\'{a}}r et al.~\cite{husar2022reduction} design pruning
strategies for multi-agent pathfinding on large instances. Cappanera et al.~\cite{caganoro23}
adopt logic-based Benders decomposition for healthcare scheduling. These
approaches rely on domain expertise and do not transfer to other problems.

Studies on generalising conflict clauses~\cite{DBLP:journals/ai/GebserKS12}
have been conducted for ASP
encodings~\cite{weinzierl2013learning,DBLP:conf/iclp/GebserKKL0S16,DBLP:journals/tplp/TaupeWF20}
and temporal domains~\cite{DBLP:journals/tplp/RomeroSS25}. Our approach differs
in not being limited to streamliners derived from a particular constraint type
(e.g., conflict clauses), but rather generating constraints from diverse sources.

\section{Streamlining ASP}
\label{sec:method}

\begin{figure}[t]
    \centering
    \includegraphics[width=1\linewidth]{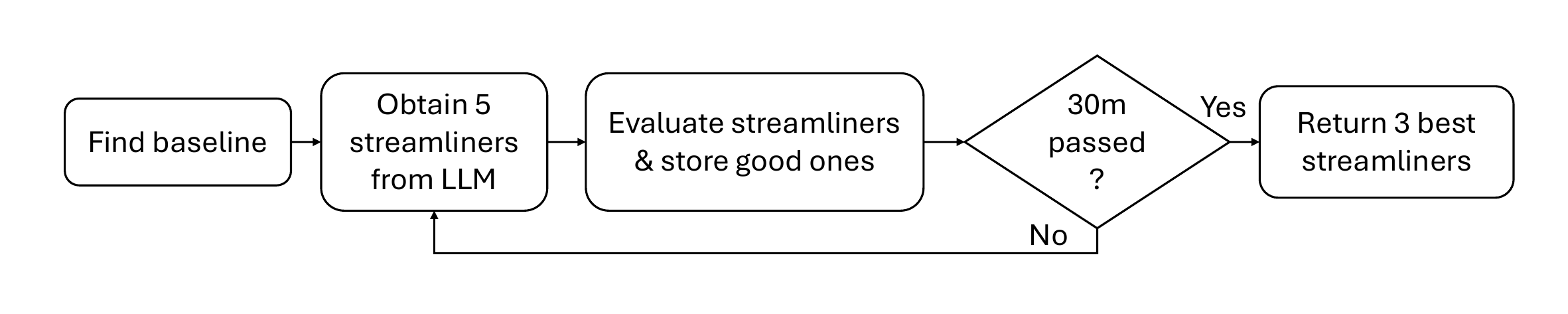}
    \caption{Fully automated pipeline of the approach}
    \label{fig:pipeline}
\end{figure}

The pipeline in Figure~\ref{fig:pipeline} illustrates our approach, which is based on StreamLLM~\cite{voboril-jair-2025}. It runs fully automated without human interaction. As input, it requires an ASP encoding together with a few satisfiable training instances that can be solved within a few seconds. First, we run the original encoding on the training instances to obtain baseline running times. Then, we provide the LLM with the encoding and the prompt in Figure~\ref{fig:prompt}. 

\definecolor{shadecolor}{gray}{0.9}
\newenvironment{promptbox}[1]{
  \begin{shaded*}
  \textbf{#1}\\
}{\end{shaded*}}

\begin{figure}[t]
\begin{framed}
\begin{promptbox}{Objective}
Analyze the given ASP problem in Clingo and suggest five additional symmetry-breaking or implied constraints to enhance the problem-solving process.
\end{promptbox}

\begin{promptbox}{Steps}\vspace*{-10pt}
  \begin{enumerate}
    \item Analyze Content: Read the provided ASP encoding. Understand the problem being addressed, including its facts, predicates, rules and constraints.
    \item Generate additional constraints: Based on your analysis, create five unique constraints. These should eliminate equivalent solutions in a search space to reduce redundancy and improve solver efficiency.
    \item Always return your constraints as a JSON object, adhering to the structure: \{``constraint\_1'': ``\textless Clingo constraint\textgreater'', ..., ``constraint\_5'': ``\textless Clingo constraint\textgreater''\}. Your final output should exclusively be the JSON object containing the five constraints. DO NOT include any comments.
  \end{enumerate}
\end{promptbox}

\begin{promptbox}{Compliance Rules}\vspace*{-10pt}
  \begin{enumerate}
    \item Code Quality: All Clingo code provided must be syntactically correct and functional.
    \item Creativity: You're encouraged to be innovative in proposing constraints, keeping in mind their purpose: to narrow down the search space efficiently by eliminating equivalent solutions.
  \end{enumerate}
\end{promptbox}\vspace*{-3pt}
\end{framed}
\caption{Used prompt}
\label{fig:prompt}
\end{figure}

The prompt is divided into three components: first, it states the objective, namely, to generate five additional \textsc{clingo} constraints that improve solving efficiency by eliminating symmetries or redundant solutions. Second, it provides a step-by-step description of how the LLM should analyze the encoding and derive suitable constraints. This procedural guidance follows the principles of the Chain-of-Thought technique~\cite{WeiEtal22}. Finally, the prompt specifies compliance rules that require syntactic correctness of the generated \textsc{clingo} code and encourage diversity and creativity in the proposed constraints. This prompt is similar to the one used for the original StreamLLM approach. The main differences are the following: 

\begin{itemize}
    \item Our approach asks for \textsc{clingo} constraints, while the original approach uses MiniZinc.
    \item We only ask for well-studied streamliners in ASP, namely symmetry-breaking and implied constraints. In contrast, the original StreamLLM approach is also interested in streamliners that might make some instances unsatisfiable.
    \item The prompt for the original StreamLLM approach tells the LLM how to deal with feedback on previously provided constraints. As the experiments by Voboril et al.~\cite{voboril-jair-2025} showed that this adaptive variant is not working significantly better than a variant without feedback, we omit this part.
\end{itemize}
Typically, the LLM returns single constraints. However, it is also possible that it returns multiple constraints or new rules. In our procedure, we treat them the same as a single constraint.

The obtained streamliners are individually added to the original encoding and tested on the training instances independently. A streamliner that leads to syntax errors, unsatisfiability, or has a longer running time than the original encoding on all training instances is discarded. The remaining streamliners that improve the running time on at least one training instance are stored. We repeat this procedure for 30 minutes. After that, the pipeline stops and returns the three streamliners that form the best combination under a VBE criterion on the training instances. Specifically, for each training instance, we consider the minimum running time across the original encoding and the selected streamliners, and the chosen triple minimizes the sum of these per-instance minimum running times.


The selected streamliners can then be applied to larger instances. For our experiments, we individually run the three streamlined encodings as well as 
the original 
encoding with all three streamliners, 
and we report the VBE, i.e., the fastest running time achieved by the original encoding or any of the streamlined variants on each instance. This allows us to analyze the extent to which different streamliners provide complementary performance improvements.

\section{Experiments}
\label{sec:experiments}

The experiments were conducted on compute nodes with 2.40GHz, 10-core 2×Intel Xeon E5-2640~v4 processors. We used \textsc{clingo} 5.8.0 as the ASP solver, which was invoked via its Python API with Python 3.11.5. In every iteration, we randomly decided for one of the following LLMs: Claude 4.5 Sonnet (anthropic/claude-sonnet-4.5), GPT-5 Mini (openai/gpt-5-mini), Gemini 3 Pro (google/gemini-3-pro-preview), Mistral Large (mistralai/mistral-large-2512), and Deepseek V3.2 (deepseek/deepseek-v3.2). All models were accessed via the OpenRouter API using identical prompts and generation parameters to ensure comparability across runs. Our repository~\cite{voboril-2026-18378760} provides the original and streamlined encodings, instances, and a table of running times for each problem, together with the code used for our approach.

For testing our method, we investigate its application on three combinatorial problems derived from the 2011 ASP competition \cite{calimeri2011third}, namely,   
 \textit{Partner Units Problem} (PUP), \textit{Sokoban} and \textit{Towers of Hanoi}, focusing on the decision version of these problems. 
Our results compare three individually streamlined encodings, their combined version, the original encoding, and a VBE that picks the fastest encoding per instance.
For each problem, we did two runs to examine how much the streamliners generated by our approach vary.
To account for unsolved instances, we aggregate running times using penalized average running times. Specifically, timeouts are penalized by a factor of 2 (PAR2), where a timeout corresponds to 20 minutes.
Instances on which all encodings led to timeouts are excluded from the aggregated running times to avoid large offsets due to their PAR2 times.
In the following subsections, we present and discuss the results for each problem and overall.

\subsection{Partner Units Problem}\label{subsec:pup}
\begin{lstlisting}[float=tbp,label=lst:pup,caption={PUP Streamliners}]
% Constraint P1
:- comUnit(U), #count{Z: unit2zone(U,Z)} = 0, #count{S: unit2sensor(U,S)} > 0.

% Constraint P2
:- unit2sensor(U1,S), unit2sensor(U2,S), U1 > U2.

% Constraint P3
:- unit2zone(U+1, _), not unit2zone(U, _), comUnit(U), comUnit(U+1).

% Constraint P4
:- comUnit(U), unit2zone(U,Z1), unit2zone(U,Z2), Z1 < Z2, 
   #count{S: zone2sensor(Z1,S), zone2sensor(Z2,S)} = 0.

% Constraint P5
:- unit2sensor(U1,S), unit2sensor(U2,S), U1 > U2.

% Constraint P6
:- comUnit(U1), comUnit(U2), U1 < U2, sensor(S), 
   S = #min{S': sensor(S')}, unit2sensor(U2,S), #count{Z: unit2zone(U1,Z)} = 0.
\end{lstlisting}

The Partner Units Problem (PUP) is an abstract representation of configuration problems occurring in railway safety or building security systems \cite{DBLP:journals/jcss/TeppanFG16}. Its input consists of a set $U$ of units and a bipartite graph $G=(S,Z,E)$, where $S$ is a set of sensors, $Z$ is a set of security/safety zones, and $E$ is a relation between $S$ and $Z$. The task is to find an assignment of vertices $v \in S\cup Z$ to units $U$ such that the following conditions hold for each unit and integers  $\ucap$ and $\iucap$:
(i) each unit contains at most $\ucap$ many sensors and $\ucap$ many zones; and
(ii) each unit has at most $\iucap$ adjacent units, where the units $u_1\neq u_2$ are adjacent whenever $v_i \in u_1$ and $v_j \in u_2$ for some $(v_i, v_j) \in E$.
For this problem, we used the benchmark suite by Aschinger et al.~\cite{DBLP:conf/cpaior/AschingerDFGJRT11}, focusing on the \textit{double}
instance collections with $\ucap=\iucap=2$, of which we extracted $7$ instances for training and $26$ instances for testing.  

Listing~\ref{lst:pup} shows the streamliners that we obtained with our approach in two runs, the former producing constraints P1, P2, and P3 and the latter producing constraints P4, P5, and P6.
Note that the constraints P2 and P5, obtained in separate runs, coincide and are redundant in the sense that they reassert a condition already implied by the given encoding: no sensor can be assigned to multiple units.
The constraints P3 and P6 break symmetries on the assignment of sensors and zones to units, without impairing satisfiability because the identifiers/integers for the units in $U$ are interchangeable.
Unlike that, the constraints P1 and P4 impose the non-trivial conditions that each unit must contain some zone or that two zones belonging to the same unit must have some sensor in common, respectively.
These two streamliners preserve the satisfiability of our training instances (and the test instances as well), while their general validity is not obvious and would necessitate a formal proof.

\begin{figure}[tbp]
    \centering
    \begin{subfigure}{0.49\textwidth}
        \centering
        \includegraphics[width=\linewidth]{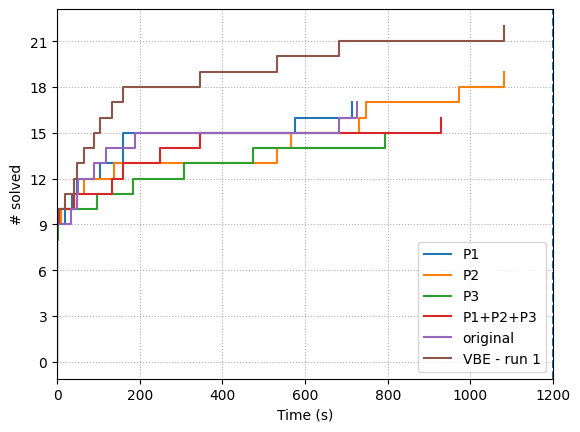}
    \end{subfigure}
    \hfill
    \begin{subfigure}{0.49\textwidth}
        \centering
        \includegraphics[width=\linewidth]{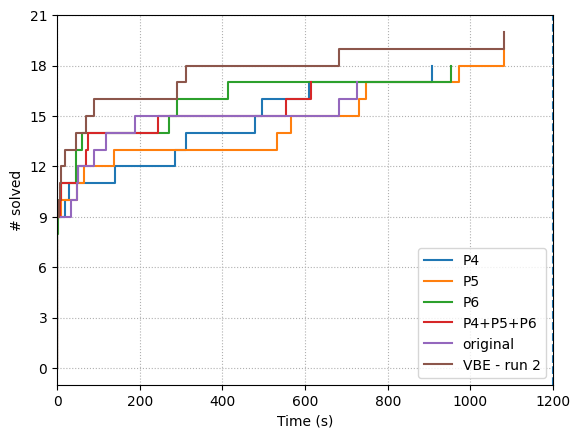}
    \end{subfigure}
    \caption{Results for PUP for run 1 (left) and run 2 (right)}
    \label{fig:cactus-pup}
\end{figure}

Figure~\ref{fig:cactus-pup} displays two cactus plots showing the performance obtained on the test set by the constraints resulting from two runs of our approach. Each plot shows the number of instances solved by $(i)$ the baseline encoding (original),  $(ii)$ the baseline encoding combined with each single streamliner constraint, $(iii)$ the baseline encoding combined with all three constraints produced in the same run, and $(iv)$ the VBE, i.e., the best-performing result per instance across all considered encodings. While the VBE could solve $22$ resp.\ $20$ instances, the original encoding could only solve $17$ instances within the time limit. 
This indicates that different streamlined encodings contribute complementary strengths, as no single encoding strictly dominates across all instances.
Table~\ref{tab:results-pup} shows the accumulated running times for the two runs,
using PAR2 for timeouts. In particular, we observe that most constraints improve the accumulated running time on the test instances. Only P3 performed worse overall than the original encoding, since it is effective on a few test instances only. The VBE indicates a potential running time reduction by 76\% (resp.\ 47\% in run~2).
Out of the $26$ test instances, four instances could not be solved with any encoding and we exclude their PAR2 times in Table~\ref{tab:results-pup}.

\begin{table}[tbp]
\caption{Sum of aggregated running times in seconds for all PUP instances}
\centering
\label{tab:results-pup}
\begin{tabular}{ccccccc}
\toprule
&  \textbf{P1} & \textbf{P2} & \textbf{P3} & \textbf{P1+P2+P3} & \textbf{original} & \textbf{VBE}\\
\emph{run 1} & 13,813 & 12,091 & 18,660 & 16,265 & 13,938 & 3,302\\
\midrule
&  \textbf{P4} & \textbf{P5} & \textbf{P6} & \textbf{P4+P5+P6} & \textbf{original} & \textbf{VBE}\\
\emph{run 2} & 12,880 & 12,091 & 11,696 & 13,627 & 13,938 & 7,412\\
\bottomrule
\end{tabular}
\end{table}

\paragraph{Lifting Symmetry Breaking with Inductive Logic Programming.}
The work of Tarzariol et al.~\cite{tagesc22b} has a similar target to our approach, namely, to learn first-order constraints preserving the satisfiability of instances. 
In particular, one of the problems tackled is PUP, where the instances of the family \textit{double} showed more difficulty, specifically on satisfiable instances. The constraints obtained in the work of Tarzariol et al.~\cite{tagesc22b} outperformed our VBE for the first $15$ test instances, yielding an improvement of $87\%$ over the original encoding, whereas the VBE achieved improvements of $64\%$ (resp.\ $62\%$ for the second run). However, they did not manage to return any solution within the considered timeout for the remaining $7$ instances. When aggregating running times over all instances using PAR2, the approach performs $22\%$ worse than the original encoding. It is worth noting that our constraints were generated by LLMs without specifying any restrictions on their structure, whereas for the work of Tarzariol et al.~\cite{tagesc22b}, a language bias must be provided. As a result, the constraints we learned have a richer structure, e.g., aggregates, arithmetic expressions such as the increment of numerical variables, 
built-in comparison operators like ``\lstinline{<}'', etc.


\subsection{Sokoban}
Sokoban is a planning problem, where each instance defines the (grid) layout of a warehouse, consisting of walls and locations (possibly designed for storage) that can hold at most one box each, and a starting situation describing the position of the agent, called sokoban, as well as boxes in the warehouse. 
The sokoban can walk on locations (unless occupied by some box) and push single boxes onto unoccupied locations. A solution of the problem is a sequence of actions to move all boxes onto storage locations.
For this problem, we split the benchmark suite by Bomanson et al.~\cite{bogejakasc16a} into $10$ training instances and $18$ test instances.

\begin{lstlisting}[float=tbp,label=lst:sokoban,caption={Sokoban Streamliners}]
% Constraint S1
:- push(X1,Y1,X2,Y2,DX,DY,T), push(X3,Y3,X4,Y4,DX,DY,T), X1 < X3.

% Constraint S2
:- push(X1,Y1,X2,Y2,DX,DY,T), push(X3,Y3,X4,Y4,DX,DY,T), X1 = X3, Y1 < Y3.

% Constraint S3
:- push(X1,Y1,X2,Y2,DX,DY,T), push(X3,Y3,X4,Y4,DX,DY,T+1), X1+Y1 > X3+Y3, 
   not box(X2,Y2,T), X2 != X3, Y2 != Y3.

% Constraint S4
:- push(X1,Y1,X2,Y2,DX,DY,T), push(X2,Y2,X1,Y1,ODX,ODY,T+1). 

% Constraint S5
:- push(X1,Y1,X2,Y2,DX,DY,T), push(X1,Y1,X3,Y3,DX,DY,T), X2 = X3, Y2 < Y3.

% Constraint S6
:- push(X1,Y1,X2,Y2,1,0,T), push(X3,Y3,X4,Y4,-1,0,T), X2 = X4, Y2 < Y4.
\end{lstlisting}

The streamliners obtained in two separate runs, denoted by S1, S2, and S3 or S4, S5, and S6, respectively, are shown in Listing~\ref{lst:sokoban}.
The majority of them, namely, S1, S2, S5, and S6, redundantly assert that different kinds of push actions cannot be performed simultaneously.
Similarly, the condition expressed by S3, i.e., the target of a push action must contain a box in the next state, is redundant, where also taking the next push action (at time \lstinline{T+1}) into account is actually unnecessary.
While the constraint S4, which states that a push action should not be immediately reversed, may seem straightforward too, it amounts to a satisfiability-preserving policy rather than a strictly necessary condition.

\begin{figure}[tbp]
    \centering
    \begin{subfigure}{0.49\textwidth}
        \centering
        \includegraphics[width=\linewidth]{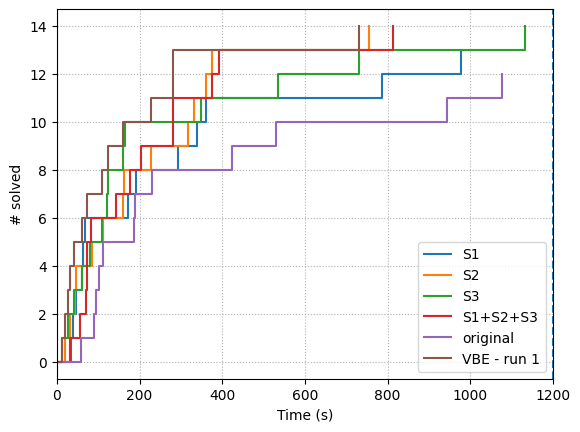}
    \end{subfigure}
    \hfill
    \begin{subfigure}{0.49\textwidth}
        \centering
        \includegraphics[width=\linewidth]{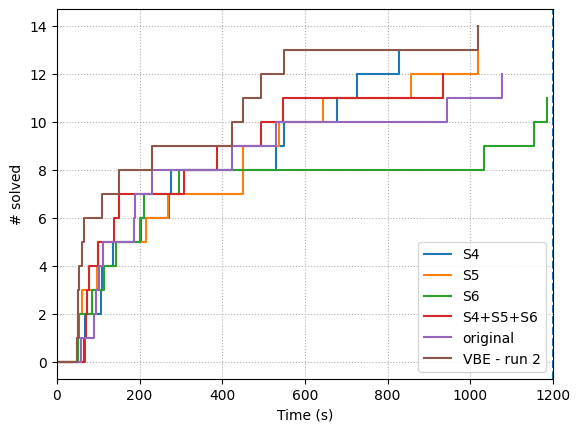}
    \end{subfigure}
    \caption{Results for Sokoban for run 1 (left) and run 2 (right)}
    \label{fig:cactus-soko}
\end{figure}

Similar to PUP in the previous subsection, Figure~\ref{fig:cactus-soko} provides two cactus plots showing the performance obtained on the test set by the constraints resulting from two runs of our approach. 
The original encoding could solve $12$ instances within the timeout, while we managed to solve $2$ more instances when considering S2, S3, or the combination of all three constraints from the first run. 
Regarding the second run, only the VBE solved $14$ instances. 
Table~\ref{tab:results-soko} shows the accumulated running times in seconds for the instances solved with at least one 
encoding, thus leaving out four of the $18$ test instances.  
In particular, the table shows that the VBE in the first run achieves a 75\% reduction in running time compared to the original encoding, while it yields a 57\% reduction in the second run.

\begin{table}[tbp]
\centering
\caption{Sum of running times in seconds for all Sokoban instances}
\label{tab:results-soko}
\begin{tabular}{ccccccc}
\toprule
& \textbf{S1} & \textbf{S2} & \textbf{S3} & \textbf{S1+S2+S3} & \textbf{original} & \textbf{VBE}\\
\emph{run 1} & 5,831 & 3,026 & 3,648 & 3,051 & 8,837 & 2,178\\
\midrule
& \textbf{S4} & \textbf{S5} & \textbf{S6} & \textbf{S4+S5+S6} & \textbf{original} & \textbf{VBE}\\
\emph{run 2} & 6,948 & 7,197 & 11,727 & 8,144 & 8,837 & 3,758\\
\bottomrule
\end{tabular}
\end{table}

\subsection{Towers of Hanoi}
Towers of Hanoi is a puzzle where the input is the initial configuration of $n$ disks of different sizes on $3$ pegs. Usually, in the initial configuration, the disks are all placed on the first peg. 
The goal is to find a sequence of movements of the disks such that all disks are moved from the first peg to the third one. Only the topmost disk of a peg can be moved at a time, and a disk cannot be moved to a peg already containing some disk of smaller size.
We synthetically generated $11$ trivial instances for training, while $46$ test instances were taken from the work of Gebser et al.~\cite{DBLP:journals/tplp/GebserKKS19}.

\begin{lstlisting}[float=tbp,label=lst:hanoi,caption={Towers of Hanoi Streamliners}]
% Constraint T1
:- moved(D1,T-1), moved(D2,T), D2 <= D1, disk(D1+1).

% Constraint T2
stable(1,P,T) :- on(1,P,T), goal_on(1,P), time(T). 
stable(D,P,T) :- on(D,P,T), goal_on(D,P), stable(D-1,P,T), disk(D), disk(D-1),
                 time(T). 
:- move(D,P,T), stable(D,P_prev,T-1), on(D,P_prev,T-1).

% Constraint T3: a comment

% Constraint T4
:- disk(D), move(D,P1,T1), move(D,P2,T2), move(D,P1,T3), T1 < T2, T2 < T3, 
   T3 = T2 + 1, P1 = P2.

% Constraint T5
:- move(D1,P1,T), move(D2,P2,T+1), D1 < D2, on(D1,Px,T-1), on(D2,Py,T), 
   Px = Py, P1 != Py, P2 = Px.

% Constraint T6
:- time(T), T > 2, move(D,P1,T-1), move(D,P2,T-2), P1 = P2, move(D,P1,T).
\end{lstlisting}

Listing~\ref{lst:hanoi} presents the streamliners T1, T2, and T3, obtained in the first run, and the streamliners T4, T5, and T6 from the second run. 
Notably, T3 is not an actual constraint but merely a comment (thus violating the instructions of Step 3 in Figure~\ref{fig:prompt}). Adding such a comment does not affect the solving performance. 
However, due to minor measurement noise in the running times of the training instances, selecting T1 and T2 (along with a comment T3) was apparently better for the performance than selecting a third proper constraint.
While the constraints T4, T5, and T6 seem complex, all three reject redundant null moves placing a disk back on its current peg, which is already prohibited by the encoding because the current peg is not a valid target for moving a disk.
The constraint T1 is also redundant yet less straightforward by asserting the implied condition that a disk cannot be moved when the two other pegs are already occupied by smaller disks.%
\footnote{If \lstinline{D1 = D2}, 
T1 asserts the policy that a disk should not be moved and immediately placed back on its previous peg.}
Interestingly, T2 constitutes a subprogram defining the auxiliary predicate
\lstinline{stable} to identify disks that have reached their goal configuration.
Along with a constraint rejecting further moves of such disks, the streamliner T2 imposes the policy to preserve already established parts of the goal configuration.%
\footnote{Similar ideas have been incorporated, e.g., in Blocksworld planning encodings \cite{gekakasc12a}, but we are unaware of any Towers of Hanoi encoding in the ASP literature from which the LLM could have directly retrieved T2.}

\begin{figure}[btp]
    \centering
    \begin{subfigure}{0.49\textwidth}
        \centering
        \includegraphics[width=\linewidth]{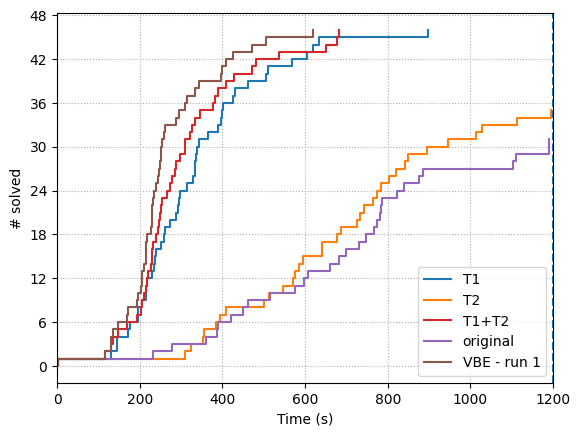}
    \end{subfigure}
    \hfill
    \begin{subfigure}{0.49\textwidth}
        \centering
        \includegraphics[width=\linewidth]{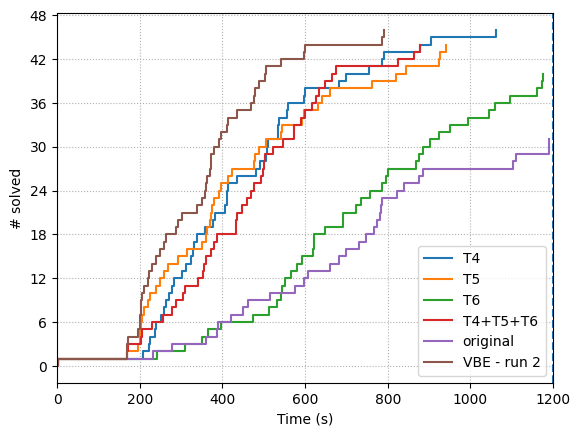}
    \end{subfigure}
    \caption{Results for Towers of Hanoi for run 1 (left) and run 2 (right)}
    \label{fig:cactus-toh}
\end{figure}

The cactus plot in Figure~\ref{fig:cactus-toh} and the accumulated running times in Table~\ref{tab:results-toh} clearly show that the original encoding performed worse than all streamlined encodings and the VBE in particular. While the original encoding only solved 31 instances within the timeout, the VBE could solve all 46 test instances in both runs. Further, it could reduce the running time by 79\% resp. 71\% in the second run.

\subsection{Analysis of Streamliners}\label{subsec:discussion}

Within the 30 minutes allocated to training, our pipeline (cf.\ Figure~\ref{fig:pipeline}) performed between 10 and 29 LLM calls. The amount of LLM calls in each run primarily depends on the number and size of the training instances. 
Comparing the generated streamliners, 52\% resulted in performance improvements on the training instances. In contrast, 19\% produced syntax errors, 9\% led to unsatisfiability on all training instances, and a further 8\% rendered some (but not all) training instances unsatisfiable. The remaining 12\% neither caused unsatisfiability nor improved upon the baseline performance.

While the selected streamliners empirically improve the performance of \textsc{clingo}, at least on the training instances, we find that they often have quadratic grounding size or can be strengthened by dropping unnecessary side conditions.
The PUP constraint P2 is an example of quadratic grounding size, as it is formulated over pairs \lstinline{U1 > U2} of units.
Its manual reformulation, denoted by P2r, achieves linear grounding size by using a \lstinline{#count} aggregate to express the same condition as P2.
Side conditions are, e.g., dropped from the Sokoban constraint S3, where
omitting the reference to a second push action yields the stronger, manually revised constraint S3r.
Our manually revised versions of generated streamliners, provided whenever the grounding size or side conditions can be reduced,
are detailed in our repository~\cite{voboril-2026-18378760}.

\begin{table}[t]
\centering
\caption{Sum of running times in seconds for all Towers of Hanoi instances}
\label{tab:results-toh}
\begin{tabular}{ccccccc}
\toprule
&  \textbf{T1} & \textbf{T2} & \textbf{T3} & \textbf{T1+T2} & \textbf{original} & \textbf{VBE}\\
\emph{run 1} & 15,108 & 49,494 & - & 13,702 & 56,710 & 11,833 \\
\midrule
&  \textbf{T4} & \textbf{T5} & \textbf{T6} & \textbf{T4+T5+T6} & \textbf{original} & \textbf{VBE}\\
\emph{run 2} & 20,853 & 23,832 & 42,569 & 24,755 & 56,710 & 16,053 \\
\bottomrule
\end{tabular}
\end{table}

Table~\ref{tab:results-pup-revised} contrasts the sum of running times in seconds obtained with the original and our manually revised constraints.
Although the revised constraints should have generally better scaling behavior, we observe a deterioration of \textsc{clingo}'s running times in almost all cases.
This comes unexpectedly and shows that the StreamLLM pipeline effectively selects streamliners that empirically improve the solver performance on the training and then also the test instances.
On the one hand, \lstinline{#count} aggregates in our grounding-oriented revisions can be more demanding to propagate than (a quadratic number of) ``flat'' constraints.
On the other hand, adding redundant constraints that are also implied by a problem encoding may negatively affect the solver's internal heuristics, so that strengthening them does not necessarily provide better performance.
From the negative results for our manually revised constraints, we conclude that the empirical approach of the StreamLLM pipeline manages to tune the solver performance better than general scaling considerations.

\begin{table}[t]
\centering
\caption{Comparison of original and revised constraints for PUP, Sokoban, and Towers of Hanoi} 
\label{tab:results-pup-revised}
\begin{tabular}{@{}l@{}c@{\hspace{8pt}}c@{\hspace{8pt}}c@{\hspace{8pt}}c@{\hspace{8pt}}c@{\hspace{8pt}}c@{\hspace{8pt}}c@{\hspace{8pt}}c@{\hspace{8pt}}c@{\hspace{8pt}}c@{\hspace{5pt}}}
\toprule
&  
\textbf{P1} & \textbf{P1r} & \textbf{P2} & \textbf{P2r} & \textbf{P6} & \textbf{P6r}\\
\emph{PUP} & 13,813 \quad & 13,810 \quad & 12,091 \quad & 13,862\quad  &
11,696 \quad & 12,077\\
\midrule
&  \textbf{S1} & \textbf{S1r} & \textbf{S2} & \textbf{S2r} & \textbf{S3} & 
\textbf{S3r} & \textbf{S5} & \textbf{S5r} & \textbf{S6} & \textbf{S6r}\\
\emph{Sokoban} & 
5,831 & 13,578 & 3,026 & 10,101 & 3,648 & 9,010 & 7,197 & 9,532 & 11,727 & 10,422
\\\midrule
& \textbf{T1} & \textbf{T1r} & \textbf{T2} & \textbf{T2r} & \textbf{T4} & \textbf{T4r} & \textbf{T5} & \textbf{T5r} & \textbf{T6} & \textbf{T6r}
\\
\emph{Hanoi}  &
15,108   & 34,236   & 49,494   & 56,211   & 20,853   & 52,795   & 23,832   & 57,188   & 42,569   & 52,795
\\\bottomrule
\end{tabular}
\end{table}

\section{Conclusion}
\label{sec:conclusion}

This paper presented the first approach of LLM-based streamliner generation for ASP. Overall, the results demonstrate that our approach is a viable and powerful tool for improving ASP solving performance. We showed that LLMs can propose syntactically valid and semantically meaningful constraints that effectively prune the search space. By filtering candidate streamliners and selecting the best-performing ones, we obtain complementary constraints. Evaluating the VBE yields runtime reductions of $47\%$-$79\%$, with most cases exceeding $70\%$, indicating substantial potential speedups when selecting the best encoding per instance. Moreover, in line with earlier findings by Voboril et al.~\cite{voboril-jair-2025}, our results suggest that only a small number of training instances is sufficient to guide effective streamliner generation, which is particularly encouraging for ASP where large curated training sets are costly. Several directions for future work remain. First, our approach should be evaluated on a broader range of ASP benchmarks, including larger instance families and additional domains, to better understand its generality and limits. Second, the quality of generated streamliners may be improved through model adaptation techniques such as fine‑tuning or knowledge distillation, as well as more advanced prompt‑optimization methods (e.g., by prompt learning or a combination with symbolic methods). Finally, extending the method to optimization problems, in line with recent work by Voboril et al.~\cite{voboril-cp-2025}, is a natural and important next step for ASP as well. 

\section{Acknowledgment}
This research was funded by the Austrian Science Fund (FWF) projects 10.55776/PAT1599524, 10.55776/\linebreak[0]{}COE12 and 10.55776/P36420.

\bibliographystyle{eptcs}
\bibliography{generic}

\begin{thebibliography}{10}
\providecommand{\bibitemdeclare}[2]{}
\providecommand{\surnamestart}{}
\providecommand{\surnameend}{}
\providecommand{\urlprefix}{Available at }
\providecommand{\url}[1]{\texttt{#1}}
\providecommand{\href}[2]{\texttt{#2}}
\providecommand{\urlalt}[2]{\href{#1}{#2}}
\providecommand{\doi}[1]{doi:\urlalt{https://doi.org/#1}{#1}}
\providecommand{\eprint}[1]{arXiv:\urlalt{https://arxiv.org/abs/#1}{#1}}
\providecommand{\bibinfo}[2]{#2}

\bibitemdeclare{inproceedings}{DBLP:conf/cpaior/AschingerDFGJRT11}
\bibitem{DBLP:conf/cpaior/AschingerDFGJRT11}
\bibinfo{author}{Markus \surnamestart Aschinger\surnameend},
  \bibinfo{author}{Conrad \surnamestart Drescher\surnameend},
  \bibinfo{author}{Gerhard \surnamestart Friedrich\surnameend},
  \bibinfo{author}{Georg \surnamestart Gottlob\surnameend},
  \bibinfo{author}{Peter \surnamestart Jeavons\surnameend},
  \bibinfo{author}{Anna \surnamestart Ryabokon\surnameend} \&
  \bibinfo{author}{Evgenij \surnamestart Thorstensen\surnameend}
  (\bibinfo{year}{2011}): \emph{\bibinfo{title}{Optimization Methods for the
  Partner Units Problem}}.
\newblock In: {\slshape \bibinfo{booktitle}{Proceedings of the 8th
  International Conference on Integration of {AI} and {OR} Techniques in
  Constraint Programming for Combinatorial Optimization Problems, {CPAIOR}
  2011}}, {\slshape \bibinfo{series}{Lecture Notes in Computer Science}}
  \bibinfo{volume}{6697}, \bibinfo{publisher}{Springer}, pp.
  \bibinfo{pages}{4--19}, \doi{10.1007/978-3-642-21311-3_4}.

\bibitemdeclare{article}{bogejakasc16a}
\bibitem{bogejakasc16a}
\bibinfo{author}{Jori \surnamestart Bomanson\surnameend},
  \bibinfo{author}{Martin \surnamestart Gebser\surnameend},
  \bibinfo{author}{Tomi \surnamestart Janhunen\surnameend},
  \bibinfo{author}{Benjamin \surnamestart Kaufmann\surnameend} \&
  \bibinfo{author}{Torsten \surnamestart Schaub\surnameend}
  (\bibinfo{year}{2016}): \emph{\bibinfo{title}{Answer Set Programming Modulo
  Acyclicity}}.
\newblock {\slshape \bibinfo{journal}{Fundamenta Informaticae}}
  \bibinfo{volume}{147}(\bibinfo{number}{1}), pp. \bibinfo{pages}{63--91},
  \doi{10.3233/FI-2016-1398}.

\bibitemdeclare{article}{CalimeriFGIKKLM20}
\bibitem{CalimeriFGIKKLM20}
\bibinfo{author}{Francesco \surnamestart Calimeri\surnameend},
  \bibinfo{author}{Wolfgang \surnamestart Faber\surnameend},
  \bibinfo{author}{Martin \surnamestart Gebser\surnameend},
  \bibinfo{author}{Giovambattista \surnamestart Ianni\surnameend},
  \bibinfo{author}{Roland \surnamestart Kaminski\surnameend},
  \bibinfo{author}{Thomas \surnamestart Krennwallner\surnameend},
  \bibinfo{author}{Nicola \surnamestart Leone\surnameend},
  \bibinfo{author}{Marco \surnamestart Maratea\surnameend},
  \bibinfo{author}{Francesco \surnamestart Ricca\surnameend} \&
  \bibinfo{author}{Torsten \surnamestart Schaub\surnameend}
  (\bibinfo{year}{2020}): \emph{\bibinfo{title}{{ASP}-Core-2 Input Language
  Format}}.
\newblock {\slshape \bibinfo{journal}{Theory and Practice of Logic
  Programming}} \bibinfo{volume}{20}(\bibinfo{number}{2}), pp.
  \bibinfo{pages}{294--309}, \doi{10.1017/S1471068419000450}.

\bibitemdeclare{article}{calimeri2011third}
\bibitem{calimeri2011third}
\bibinfo{author}{Francesco \surnamestart Calimeri\surnameend},
  \bibinfo{author}{Giovambattista \surnamestart Ianni\surnameend} \&
  \bibinfo{author}{Francesco \surnamestart Ricca\surnameend}
  (\bibinfo{year}{2014}): \emph{\bibinfo{title}{The Third Open Answer Set
  Programming Competition}}.
\newblock {\slshape \bibinfo{journal}{Theory and Practice of Logic
  Programming}} \bibinfo{volume}{14}(\bibinfo{number}{1}), pp.
  \bibinfo{pages}{117--135}, \doi{10.1017/S1471068412000105}.

\bibitemdeclare{article}{caganoro23}
\bibitem{caganoro23}
\bibinfo{author}{Paola \surnamestart Cappanera\surnameend},
  \bibinfo{author}{Marco \surnamestart Gavanelli\surnameend},
  \bibinfo{author}{Maddalena \surnamestart Nonato\surnameend} \&
  \bibinfo{author}{Marco \surnamestart Roma\surnameend} (\bibinfo{year}{2023}):
  \emph{\bibinfo{title}{Logic-Based Benders Decomposition in Answer Set
  Programming for Chronic Outpatients Scheduling}}.
\newblock {\slshape \bibinfo{journal}{Theory and Practice of Logic
  Programming}} \bibinfo{volume}{23}(\bibinfo{number}{4}), pp.
  \bibinfo{pages}{848--864}, \doi{10.1017/s147106842300025x}.

\bibitemdeclare{inproceedings}{CoppolilloEtAl2024}
\bibitem{CoppolilloEtAl2024}
\bibinfo{author}{Erica \surnamestart Coppolillo\surnameend},
  \bibinfo{author}{Francesco \surnamestart Calimeri\surnameend},
  \bibinfo{author}{Giuseppe \surnamestart Manco\surnameend},
  \bibinfo{author}{Simona \surnamestart Perri\surnameend} \&
  \bibinfo{author}{Francesco \surnamestart Ricca\surnameend}
  (\bibinfo{year}{2024}): \emph{\bibinfo{title}{{LLASP}: Fine-Tuning Large
  Language Models for Answer Set Programming}}.
\newblock In: {\slshape \bibinfo{booktitle}{Proceedings of the 21st
  International Conference on Principles of Knowledge Representation and
  Reasoning, {KR} 2024}}, \bibinfo{publisher}{ijcai.org},
  \doi{10.24963/kr.2024/78}.

\bibitemdeclare{article}{debo2016a}
\bibitem{debo2016a}
\bibinfo{author}{Jo~\surnamestart Devriendt\surnameend} \&
  \bibinfo{author}{Bart \surnamestart Bogaerts\surnameend}
  (\bibinfo{year}{2016}): \emph{\bibinfo{title}{{BreakID}: Static Symmetry
  Breaking for {ASP} (System Description)}}.
\newblock {\slshape \bibinfo{journal}{CoRR}} \bibinfo{volume}{abs/1608.08447},
  \doi{10.48550/arXiv.1608.08447}.

\bibitemdeclare{article}{debobrde16a}
\bibitem{debobrde16a}
\bibinfo{author}{Jo~\surnamestart Devriendt\surnameend}, \bibinfo{author}{Bart
  \surnamestart Bogaerts\surnameend}, \bibinfo{author}{Maurice \surnamestart
  Bruynooghe\surnameend} \& \bibinfo{author}{Marc \surnamestart
  Denecker\surnameend} (\bibinfo{year}{2016}): \emph{\bibinfo{title}{On Local
  Domain Symmetry for Model Expansion}}.
\newblock {\slshape \bibinfo{journal}{Theory and Practice of Logic
  Programming}} \bibinfo{volume}{16}(\bibinfo{number}{5-6}), pp.
  \bibinfo{pages}{636--652}, \doi{10.1017/S1471068416000508}.

\bibitemdeclare{article}{drtiwa11a}
\bibitem{drtiwa11a}
\bibinfo{author}{Christian \surnamestart Drescher\surnameend},
  \bibinfo{author}{Oana \surnamestart Tifrea\surnameend} \&
  \bibinfo{author}{Toby \surnamestart Walsh\surnameend} (\bibinfo{year}{2011}):
  \emph{\bibinfo{title}{Symmetry-Breaking Answer Set Solving}}.
\newblock {\slshape \bibinfo{journal}{AI Communications}}
  \bibinfo{volume}{24}(\bibinfo{number}{2}), pp. \bibinfo{pages}{177--194},
  \doi{10.3233/AIC-2011-0495}.

\bibitemdeclare{article}{gehakalisc15a}
\bibitem{gehakalisc15a}
\bibinfo{author}{Martin \surnamestart Gebser\surnameend},
  \bibinfo{author}{Amelia \surnamestart Harrison\surnameend},
  \bibinfo{author}{Roland \surnamestart Kaminski\surnameend},
  \bibinfo{author}{Vladimir \surnamestart Lifschitz\surnameend} \&
  \bibinfo{author}{Torsten \surnamestart Schaub\surnameend}
  (\bibinfo{year}{2015}): \emph{\bibinfo{title}{Abstract {G}ringo}}.
\newblock {\slshape \bibinfo{journal}{Theory and Practice of Logic
  Programming}} \bibinfo{volume}{15}(\bibinfo{number}{4-5}), pp.
  \bibinfo{pages}{449--463}, \doi{10.1017/S1471068415000150}.

\bibitemdeclare{inproceedings}{DBLP:conf/iclp/GebserKKL0S16}
\bibitem{DBLP:conf/iclp/GebserKKL0S16}
\bibinfo{author}{Martin \surnamestart Gebser\surnameend},
  \bibinfo{author}{Roland \surnamestart Kaminski\surnameend},
  \bibinfo{author}{Benjamin \surnamestart Kaufmann\surnameend},
  \bibinfo{author}{Patrick \surnamestart L{\"{u}}hne\surnameend},
  \bibinfo{author}{Javier \surnamestart Romero\surnameend} \&
  \bibinfo{author}{Torsten \surnamestart Schaub\surnameend}
  (\bibinfo{year}{2016}): \emph{\bibinfo{title}{Answer Set Solving With
  Generalized Learned Constraints}}.
\newblock In: {\slshape \bibinfo{booktitle}{Technical Communications of the
  32nd International Conference on Logic Programming, {ICLP} 2016}}, {\slshape
  \bibinfo{series}{OASIcs}}~\bibinfo{volume}{52}, \bibinfo{publisher}{Schloss
  Dagstuhl - Leibniz-Zentrum f{\"{u}}r Informatik}, pp.
  \bibinfo{pages}{9:1--9:15}, \doi{10.4230/OASIcs.ICLP.2016.9}.

\bibitemdeclare{book}{gekakasc12a}
\bibitem{gekakasc12a}
\bibinfo{author}{Martin \surnamestart Gebser\surnameend},
  \bibinfo{author}{Roland \surnamestart Kaminski\surnameend},
  \bibinfo{author}{Benjamin \surnamestart Kaufmann\surnameend} \&
  \bibinfo{author}{Torsten \surnamestart Schaub\surnameend}
  (\bibinfo{year}{2012}): \emph{\bibinfo{title}{Answer Set Solving in
  Practice}}.
\newblock \bibinfo{publisher}{Morgan and Claypool Publishers},
  \doi{10.1007/978-3-031-01561-8}.

\bibitemdeclare{article}{DBLP:journals/tplp/GebserKKS19}
\bibitem{DBLP:journals/tplp/GebserKKS19}
\bibinfo{author}{Martin \surnamestart Gebser\surnameend},
  \bibinfo{author}{Roland \surnamestart Kaminski\surnameend},
  \bibinfo{author}{Benjamin \surnamestart Kaufmann\surnameend} \&
  \bibinfo{author}{Torsten \surnamestart Schaub\surnameend}
  (\bibinfo{year}{2019}): \emph{\bibinfo{title}{Multi-Shot {ASP} Solving With
  Clingo}}.
\newblock {\slshape \bibinfo{journal}{Theory and Practice of Logic
  Programming}} \bibinfo{volume}{19}(\bibinfo{number}{1}), pp.
  \bibinfo{pages}{27--82}, \doi{10.1017/S1471068418000054}.

\bibitemdeclare{article}{DBLP:journals/ai/GebserKS12}
\bibitem{DBLP:journals/ai/GebserKS12}
\bibinfo{author}{Martin \surnamestart Gebser\surnameend},
  \bibinfo{author}{Benjamin \surnamestart Kaufmann\surnameend} \&
  \bibinfo{author}{Torsten \surnamestart Schaub\surnameend}
  (\bibinfo{year}{2012}): \emph{\bibinfo{title}{Conflict-Driven Answer Set
  Solving: From Theory to Practice}}.
\newblock {\slshape \bibinfo{journal}{Artificial Intelligence}}
  \bibinfo{volume}{187}, pp. \bibinfo{pages}{52--89},
  \doi{10.1016/j.artint.2012.04.001}.

\bibitemdeclare{inproceedings}{GomesSellmann2004}
\bibitem{GomesSellmann2004}
\bibinfo{author}{Carla~P. \surnamestart Gomes\surnameend} \&
  \bibinfo{author}{Meinolf \surnamestart Sellmann\surnameend}
  (\bibinfo{year}{2004}): \emph{\bibinfo{title}{Streamlined Constraint
  Reasoning}}.
\newblock In: {\slshape \bibinfo{booktitle}{Proceedings of the 10th
  International Conference on Principles and Practice of Constraint
  Programming, {CP} 2004}}, {\slshape \bibinfo{series}{Lecture Notes in
  Computer Science}} \bibinfo{volume}{3258}, \bibinfo{publisher}{Springer}, pp.
  \bibinfo{pages}{274--289}, \doi{10.1007/978-3-540-30201-8_22}.

\bibitemdeclare{inproceedings}{husar2022reduction}
\bibitem{husar2022reduction}
\bibinfo{author}{Matej \surnamestart Hus{\'{a}}r\surnameend},
  \bibinfo{author}{Jir{\'{\i}} \surnamestart Svancara\surnameend},
  \bibinfo{author}{Philipp \surnamestart Obermeier\surnameend},
  \bibinfo{author}{Roman \surnamestart Bart{\'{a}}k\surnameend} \&
  \bibinfo{author}{Torsten \surnamestart Schaub\surnameend}
  (\bibinfo{year}{2022}): \emph{\bibinfo{title}{Reduction-Based Solving of
  Multi-Agent Pathfinding on Large Maps Using Graph Pruning}}.
\newblock In: {\slshape \bibinfo{booktitle}{Proceedings of the 21st
  International Conference on Autonomous Agents and Multiagent Systems, {AAMAS}
  2022}}, \bibinfo{publisher}{{IFAAMAS}}, pp. \bibinfo{pages}{624--632},
  \doi{10.5555/3535850.3535921}.

\bibitemdeclare{inproceedings}{IshayYL2023}
\bibitem{IshayYL2023}
\bibinfo{author}{Adam \surnamestart Ishay\surnameend}, \bibinfo{author}{Zhun
  \surnamestart Yang\surnameend} \& \bibinfo{author}{Joohyung \surnamestart
  Lee\surnameend} (\bibinfo{year}{2023}): \emph{\bibinfo{title}{Leveraging
  Large Language Models to Generate Answer Set Programs}}.
\newblock In: {\slshape \bibinfo{booktitle}{Proceedings of the 20th
  International Conference on Principles of Knowledge Representation and
  Reasoning, {KR} 2023}}, \bibinfo{publisher}{ijcai.org}, pp.
  \bibinfo{pages}{374--383}, \doi{10.24963/kr.2023/37}.

\bibitemdeclare{article}{Law2023}
\bibitem{Law2023}
\bibinfo{author}{Mark \surnamestart Law\surnameend} (\bibinfo{year}{2023}):
  \emph{\bibinfo{title}{Conflict-Driven Inductive Logic Programming}}.
\newblock {\slshape \bibinfo{journal}{Theory and Practice of Logic
  Programming}} \bibinfo{volume}{23}(\bibinfo{number}{2}), pp.
  \bibinfo{pages}{387--414}, \doi{10.1017/S1471068422000011}.

\bibitemdeclare{inproceedings}{LeoGBW2024}
\bibitem{LeoGBW2024}
\bibinfo{author}{Kevin \surnamestart Leo\surnameend}, \bibinfo{author}{Graeme
  \surnamestart Gange\surnameend}, \bibinfo{author}{Maria~Garcia \surnamestart
  de~la Banda\surnameend} \& \bibinfo{author}{Mark \surnamestart
  Wallace\surnameend} (\bibinfo{year}{2024}): \emph{\bibinfo{title}{Automatic
  Core-Guided Reformulation via Constraint Explanation and Condition
  Learning}}.
\newblock In: {\slshape \bibinfo{booktitle}{Proceedings of the 38th {AAAI}
  Conference on Artificial Intelligence, {AAAI} 2024}},
  \bibinfo{publisher}{{AAAI} Press}, pp. \bibinfo{pages}{8065--8072},
  \doi{10.1609/aaai.v38i8.28645}.

\bibitemdeclare{article}{DBLP:journals/tplp/RomeroSS25}
\bibitem{DBLP:journals/tplp/RomeroSS25}
\bibinfo{author}{Javier \surnamestart Romero\surnameend},
  \bibinfo{author}{Torsten \surnamestart Schaub\surnameend} \&
  \bibinfo{author}{Klaus \surnamestart Strauch\surnameend}
  (\bibinfo{year}{2025}): \emph{\bibinfo{title}{On the Generalization of
  Learned Constraints for {ASP} Solving in Temporal Domains}}.
\newblock {\slshape \bibinfo{journal}{Theory and Practice of Logic
  Programming}} \bibinfo{volume}{25}(\bibinfo{number}{2}), pp.
  \bibinfo{pages}{197--224}, \doi{10.1017/s1471068424000462}.

\bibitemdeclare{incollection}{sakallah09a}
\bibitem{sakallah09a}
\bibinfo{author}{Karem~A. \surnamestart Sakallah\surnameend}
  (\bibinfo{year}{2021}): \emph{\bibinfo{title}{Symmetry and Satisfiability}}.
\newblock In: {\slshape \bibinfo{booktitle}{Handbook of Satisfiability}},
  {\slshape \bibinfo{series}{Frontiers in Artificial Intelligence and
  Applications}} \bibinfo{volume}{336}, \bibinfo{publisher}{{IOS} Press}, pp.
  \bibinfo{pages}{509--570}, \doi{10.3233/FAIA200996}.

\bibitemdeclare{inproceedings}{SantanaBCF24}
\bibitem{SantanaBCF24}
\bibinfo{author}{Manuel A.~Borroto \surnamestart Santana\surnameend},
  \bibinfo{author}{Irfan \surnamestart Kareem\surnameend} \&
  \bibinfo{author}{Francesco \surnamestart Ricca\surnameend}
  (\bibinfo{year}{2024}): \emph{\bibinfo{title}{Towards Automatic Composition
  of {ASP} Programs from Natural Language Specifications}}.
\newblock In: {\slshape \bibinfo{booktitle}{Proceedings of the 33rd
  International Joint Conference on Artificial Intelligence, {IJCAI} 2024}},
  \bibinfo{publisher}{ijcai.org}, pp. \bibinfo{pages}{6198--6206},
  \doi{10.24963/ijcai.2024/685}.

\bibitemdeclare{article}{SkryaginSL2024}
\bibitem{SkryaginSL2024}
\bibinfo{author}{Arseny \surnamestart Skryagin\surnameend},
  \bibinfo{author}{Daniel \surnamestart Ochs\surnameend},
  \bibinfo{author}{Phillip \surnamestart Deibert\surnameend},
  \bibinfo{author}{Simon \surnamestart Kohaut\surnameend},
  \bibinfo{author}{Devendra~Singh \surnamestart Dhami\surnameend} \&
  \bibinfo{author}{Kristian \surnamestart Kersting\surnameend}
  (\bibinfo{year}{2024}): \emph{\bibinfo{title}{Answer Set Networks: Casting
  Answer Set Programming into Deep Learning}}.
\newblock {\slshape \bibinfo{journal}{CoRR}} \bibinfo{volume}{abs/2412.14814},
  \doi{10.48550/arXiv.2412.14814}.

\bibitemdeclare{article}{SpracklenDAM2023}
\bibitem{SpracklenDAM2023}
\bibinfo{author}{Patrick \surnamestart Spracklen\surnameend},
  \bibinfo{author}{Nguyen \surnamestart Dang\surnameend},
  \bibinfo{author}{{\"{O}}zg{\"{u}}r \surnamestart Akg{\"{u}}n\surnameend} \&
  \bibinfo{author}{Ian \surnamestart Miguel\surnameend} (\bibinfo{year}{2023}):
  \emph{\bibinfo{title}{Automated Streamliner Portfolios for Constraint
  Satisfaction Problems}}.
\newblock {\slshape \bibinfo{journal}{Artificial Intelligence}}
  \bibinfo{volume}{319}, p. \bibinfo{pages}{103915},
  \doi{10.1016/j.artint.2023.103915}.

\bibitemdeclare{misc}{SzeiderASPBench2025}
\bibitem{SzeiderASPBench2025}
\bibinfo{author}{Stefan \surnamestart Szeider\surnameend}
  (\bibinfo{year}{2025}): \emph{\bibinfo{title}{{ASP-Bench}: Problems, Ground
  Truths, and Solutions}}, \doi{10.5281/zenodo.18062939}.

\bibitemdeclare{inproceedings}{Szeider2025mcp}
\bibitem{Szeider2025mcp}
\bibinfo{author}{Stefan \surnamestart Szeider\surnameend}
  (\bibinfo{year}{2025}): \emph{\bibinfo{title}{Bridging Language Models and
  Symbolic Solvers via the Model Context Protocol}}.
\newblock In: {\slshape \bibinfo{booktitle}{Proceedings of the 28th
  International Conference on Theory and Applications of Satisfiability
  Testing, {SAT} 2025}}, {\slshape \bibinfo{series}{LIPIcs}}
  \bibinfo{volume}{341}, \bibinfo{publisher}{Schloss Dagstuhl - Leibniz-Zentrum
  f{\"{u}}r Informatik}, pp. \bibinfo{pages}{30:1--30:12},
  \doi{10.4230/LIPIcs.SAT.2025.30}.

\bibitemdeclare{article}{tagesc22}
\bibitem{tagesc22}
\bibinfo{author}{Alice \surnamestart Tarzariol\surnameend},
  \bibinfo{author}{Martin \surnamestart Gebser\surnameend} \&
  \bibinfo{author}{Konstantin \surnamestart Schekotihin\surnameend}
  (\bibinfo{year}{2022}): \emph{\bibinfo{title}{Lifting Symmetry Breaking
  Constraints With Inductive Logic Programming}}.
\newblock {\slshape \bibinfo{journal}{Machine Learning}}
  \bibinfo{volume}{111}(\bibinfo{number}{4}), pp. \bibinfo{pages}{1303--1326},
  \doi{10.1007/s10994-022-06146-3}.

\bibitemdeclare{inproceedings}{tagesc23}
\bibitem{tagesc23}
\bibinfo{author}{Alice \surnamestart Tarzariol\surnameend},
  \bibinfo{author}{Martin \surnamestart Gebser\surnameend},
  \bibinfo{author}{Konstantin \surnamestart Schekotihin\surnameend} \&
  \bibinfo{author}{Mark \surnamestart Law\surnameend} (\bibinfo{year}{2023}):
  \emph{\bibinfo{title}{Learning to Break Symmetries for Efficient Optimization
  in Answer Set Programming}}.
\newblock In: {\slshape \bibinfo{booktitle}{Proceedings of the 37th {AAAI}
  Conference on Artificial Intelligence, {AAAI} 2023}},
  \bibinfo{publisher}{{AAAI} Press}, pp. \bibinfo{pages}{6541--6549},
  \doi{10.1609/aaai.v37i5.25804}.

\bibitemdeclare{article}{tagesc22b}
\bibitem{tagesc22b}
\bibinfo{author}{Alice \surnamestart Tarzariol\surnameend},
  \bibinfo{author}{Konstantin \surnamestart Schekotihin\surnameend},
  \bibinfo{author}{Martin \surnamestart Gebser\surnameend} \&
  \bibinfo{author}{Mark \surnamestart Law\surnameend} (\bibinfo{year}{2022}):
  \emph{\bibinfo{title}{Efficient Lifting of Symmetry Breaking Constraints for
  Complex Combinatorial Problems}}.
\newblock {\slshape \bibinfo{journal}{Theory and Practice of Logic
  Programming}} \bibinfo{volume}{22}(\bibinfo{number}{4}), pp.
  \bibinfo{pages}{606--622}, \doi{10.1017/S1471068422000151}.

\bibitemdeclare{article}{DBLP:journals/tplp/TaupeWF20}
\bibitem{DBLP:journals/tplp/TaupeWF20}
\bibinfo{author}{Richard \surnamestart Taupe\surnameend},
  \bibinfo{author}{Antonius \surnamestart Weinzierl\surnameend} \&
  \bibinfo{author}{Gerhard \surnamestart Friedrich\surnameend}
  (\bibinfo{year}{2020}): \emph{\bibinfo{title}{Conflict Generalisation in
  {ASP}: Learning Correct and Effective Non-Ground Constraints}}.
\newblock {\slshape \bibinfo{journal}{Theory and Practice of Logic
  Programming}} \bibinfo{volume}{20}(\bibinfo{number}{5}), pp.
  \bibinfo{pages}{799--814}, \doi{10.1017/S1471068420000368}.

\bibitemdeclare{article}{DBLP:journals/jcss/TeppanFG16}
\bibitem{DBLP:journals/jcss/TeppanFG16}
\bibinfo{author}{Erich~Christian \surnamestart Teppan\surnameend},
  \bibinfo{author}{Gerhard \surnamestart Friedrich\surnameend} \&
  \bibinfo{author}{Georg \surnamestart Gottlob\surnameend}
  (\bibinfo{year}{2016}): \emph{\bibinfo{title}{Tractability Frontiers of the
  Partner Units Configuration Problem}}.
\newblock {\slshape \bibinfo{journal}{Journal of Computer and System Sciences}}
  \bibinfo{volume}{82}(\bibinfo{number}{5}), pp. \bibinfo{pages}{739--755},
  \doi{10.1016/j.jcss.2015.12.004}.

\bibitemdeclare{misc}{voboril-2026-18378760}
\bibitem{voboril-2026-18378760}
\bibinfo{author}{Florentina \surnamestart Voboril\surnameend},
  \bibinfo{author}{Martin \surnamestart Gebser\surnameend},
  \bibinfo{author}{Stefan \surnamestart Szeider\surnameend} \&
  \bibinfo{author}{Alice \surnamestart Tarzariol\surnameend}
  (\bibinfo{year}{2026}): \emph{\bibinfo{title}{Code and Instances for the
  Paper: Streamliners for Answer Set Programming}},
  \doi{10.5281/zenodo.18378760}.

\bibitemdeclare{inproceedings}{voboril-cp-2025}
\bibitem{voboril-cp-2025}
\bibinfo{author}{Florentina \surnamestart Voboril\surnameend},
  \bibinfo{author}{Vaidyanathan~Peruvemba \surnamestart Ramaswamy\surnameend}
  \& \bibinfo{author}{Stefan \surnamestart Szeider\surnameend}
  (\bibinfo{year}{2025}): \emph{\bibinfo{title}{Balancing {L}atin Rectangles
  With {LLM}-Generated Streamliners}}.
\newblock In: {\slshape \bibinfo{booktitle}{Proceedings of the 31st
  International Conference on Principles and Practice of Constraint
  Programming, {CP} 2025}}, {\slshape \bibinfo{series}{LIPIcs}}
  \bibinfo{volume}{340}, \bibinfo{publisher}{Schloss Dagstuhl - Leibniz-Zentrum
  f{\"{u}}r Informatik}, pp. \bibinfo{pages}{36:1--36:17},
  \doi{10.4230/LIPIcs.CP.2025.36}.

\bibitemdeclare{article}{voboril-jair-2025}
\bibitem{voboril-jair-2025}
\bibinfo{author}{Florentina \surnamestart Voboril\surnameend},
  \bibinfo{author}{Vaidyanathan~Peruvemba \surnamestart Ramaswamy\surnameend}
  \& \bibinfo{author}{Stefan \surnamestart Szeider\surnameend}
  (\bibinfo{year}{2025}): \emph{\bibinfo{title}{Generating Streamlining
  Constraints With Large Language Models}}.
\newblock {\slshape \bibinfo{journal}{Journal of Artificial Intelligence
  Research}} \bibinfo{volume}{84}, pp. \bibinfo{pages}{16:1--16:20},
  \doi{10.1613/jair.1.18965}.

\bibitemdeclare{inproceedings}{Walsh2012}
\bibitem{Walsh2012}
\bibinfo{author}{Toby \surnamestart Walsh\surnameend} (\bibinfo{year}{2012}):
  \emph{\bibinfo{title}{Symmetry Breaking Constraints: Recent Results}}.
\newblock In: {\slshape \bibinfo{booktitle}{Proceedings of the 26th {AAAI}
  Conference on Artificial Intelligence, {AAAI} 2012}},
  \bibinfo{publisher}{{AAAI} Press}, \doi{10.1609/AAAI.V26I1.8437}.

\bibitemdeclare{inproceedings}{WeiEtal22}
\bibitem{WeiEtal22}
\bibinfo{author}{Jason \surnamestart Wei\surnameend}, \bibinfo{author}{Xuezhi
  \surnamestart Wang\surnameend}, \bibinfo{author}{Dale \surnamestart
  Schuurmans\surnameend}, \bibinfo{author}{Maarten \surnamestart
  Bosma\surnameend}, \bibinfo{author}{Brian \surnamestart Richter\surnameend},
  \bibinfo{author}{Fei \surnamestart Xia\surnameend},
  \bibinfo{author}{Ed~\surnamestart Chi\surnameend}, \bibinfo{author}{Quoc~V.
  \surnamestart Le\surnameend} \& \bibinfo{author}{Denny \surnamestart
  Zhou\surnameend} (\bibinfo{year}{2022}):
  \emph{\bibinfo{title}{Chain-of-Thought Prompting Elicits Reasoning in Large
  Language Models}}.
\newblock In: {\slshape \bibinfo{booktitle}{Proceedings of the 36th
  International Conference on Neural Information Processing Systems, NIPS
  2022}}, \bibinfo{publisher}{Curran Associates Inc.}, pp.
  \bibinfo{pages}{24824--24837}.

\bibitemdeclare{inproceedings}{weinzierl2013learning}
\bibitem{weinzierl2013learning}
\bibinfo{author}{Antonius \surnamestart Weinzierl\surnameend}
  (\bibinfo{year}{2013}): \emph{\bibinfo{title}{Learning Non-Ground Rules for
  Answer-Set Solving}}.
\newblock In: {\slshape \bibinfo{booktitle}{Proceedings of the 2nd Workshop on
  Grounding and Transformations for Theories With Variables, GTTV 2013}}.

\bibitemdeclare{inproceedings}{YangIL2020}
\bibitem{YangIL2020}
\bibinfo{author}{Zhun \surnamestart Yang\surnameend}, \bibinfo{author}{Adam
  \surnamestart Ishay\surnameend} \& \bibinfo{author}{Joohyung \surnamestart
  Lee\surnameend} (\bibinfo{year}{2020}): \emph{\bibinfo{title}{{NeurASP}:
  Embracing Neural Networks into Answer Set Programming}}.
\newblock In: {\slshape \bibinfo{booktitle}{Proceedings of the 29th
  International Joint Conference on Artificial Intelligence, {IJCAI} 2020}},
  \bibinfo{publisher}{ijcai.org}, pp. \bibinfo{pages}{1755--1762},
  \doi{10.24963/ijcai.2020/243}.

\bibitemdeclare{inproceedings}{yli2023pruning}
\bibitem{yli2023pruning}
\bibinfo{author}{Anssi \surnamestart Yli-Jyr{\"{a}}\surnameend},
  \bibinfo{author}{Masood~Feyzbakhsh \surnamestart Rankooh\surnameend} \&
  \bibinfo{author}{Tomi \surnamestart Janhunen\surnameend}
  (\bibinfo{year}{2023}): \emph{\bibinfo{title}{Pruning Redundancy in Answer
  Set Optimization Applied to Preventive Maintenance Scheduling}}.
\newblock In: {\slshape \bibinfo{booktitle}{Proceedings of the 25th
  International Symposium on Practical Aspects of Declarative Languages, {PADL}
  2023}}, {\slshape \bibinfo{series}{Lecture Notes in Computer Science}}
  \bibinfo{volume}{13880}, \bibinfo{publisher}{Springer}, pp.
  \bibinfo{pages}{279--294}, \doi{10.1007/978-3-031-24841-2\_18}.

\end{thebibliography}

\newpage
\appendix
\section{Encodings}\label{sec:encodings}
This section illustrates the encodings used for the three domains studied in the paper. In particular, it provides an intuition for the meaning of the predicates appearing in the streamliners returned by our approach.

\paragraph{\textbf{Partner Units Problem (PUP).}}
The baseline PUP encoding is derived from Aschinger et al.~\cite{DBLP:conf/cpaior/AschingerDFGJRT11} and is displayed in Listing~\ref{lst:encPUP}.
A PUP instance provides atoms of the form \lstinline{zone2sensor/2}, representing the bipartite input graph, and atoms of the form \lstinline{comUnit/1} providing the IDs of the (communication) units.
Morevoer, the constants $\ucap$ and $\iucap$ are defined via the atom \lstinline{maxElements(M)} with \lstinline{M}${}=\ucap$ and \lstinline{maxPU(M)} with \lstinline{M}${}=\iucap$.
The encoding then defines the assignments of units to zones via \lstinline{unit2zone/2} and units to sensors via \lstinline{unit2sensors/2}, as well as the condition for two units to be adjacent with atoms of the form \lstinline{partnerunits/2}. Asserting respective constraints guarantees that the problem conditions hold. 

\begin{lstlisting}[float=b,label=lst:encPUP,caption={Baseline Encoding for PUP}]
zone(Z) :- zone2sensor(Z,D).
sensor(D) :- zone2sensor(Z,D).

1 { unit2zone(U,Z) : comUnit(U) } 1 :- zone(Z).
1 { unit2sensor(U,S) : comUnit(U) } 1 :- sensor(S).
:- comUnit(U), maxElements(M), M+1 { unit2zone(U,Z): zone(Z) }.
:- comUnit(U), maxElements(M), M+1 { unit2sensor(U,S): sensor(S) }.
partnerunits(U,P) :- unit2zone(U,Z), zone2sensor(Z,S), unit2sensor(P,S), U!=P.
partnerunits(U,P) :- partnerunits(P,U), comUnit(U), comUnit(P).
:- comUnit(U), maxPU(M), M+1 { partnerunits(U,P): comUnit(P) }.
\end{lstlisting}

\paragraph{\textbf{Sokoban.}}
The baseline encoding for Sokoban is shown in Listing~\ref{lst:encSokoban}.
An instance provides atoms of the form \lstinline{square/2}, \lstinline{initial_box/2}, and \lstinline{target_square/2}, representing the grid locations, the initial locations of the boxes, or their goal locations, respectively. 
Moreover, the atoms \lstinline{initial_at(X1,Y1)} and \lstinline{steps(T)} specify the initial location \lstinline{(X1,Y1)} of the sokoban and the maximum number \lstinline{T} of actions.
The Sokoban encoding provides the output predicate  \lstinline{push/7}, 
where \lstinline{push(X1,Y1,X2,Y2,DX,DY,T)} means that, at time \lstinline{T}, the sokoban pushes a box placed in the location \lstinline{(X1,Y1)} to the new location \lstinline{(X2,Y2)} in direction \lstinline{(DX,DY)} (e.g., \lstinline{(1,0)} means rightwards, \lstinline{(0,1)} means upwards, etc.). 
Auxiliary predicates (e.g., \lstinline{next/6}, \lstinline{from/5}, \lstinline{reach/3}, etc.) define relations used in constraints to produce a plan that complies with the rules of the Sokoban problem. Among them, \lstinline{box(X,Y,T)} means that, at time \lstinline{T}, a box is placed in the location \lstinline{(X,Y)}.

\begin{lstlisting}[float=tb,label=lst:encSokoban,caption={Baseline Encoding for Sokoban},lastline=44]
diff(0,1;1,0;0,-1;-1,0).

next(X1,Y1,X2,Y2,DX,DY) :- square(X1,Y1), diff(DX,DY),
                           square(X2,Y2), X2 = X1+DX, Y2 = Y1+DY.
next(X2,Y2,DX,DY)       :- next(X1,Y1,X2,Y2,DX,DY).

path(X1,Y1,X2,Y2,DX,DY) :- next(X1,Y1,X2,Y2,DX,DY), next(X1,Y1,DX,DY).
path(X1,Y1,X2,Y2,DX,DY) :- path(X1,Y1,X3,Y3,DX,DY), next(X3,Y3,X2,Y2,DX,DY).

step(1..T) :- steps(T).

box(X,Y,0) :- initial_box(X,Y).

{push(X1,Y1,X2,Y2,DX,DY,T) : path(X1,Y1,X2,Y2,DX,DY)} 1 :- step(T).

from(X1,Y1,DX,DY,T) :- push(X1,Y1,X2,Y2,DX,DY,T).
from(X1,Y1,T)       :- from(X1,Y1,DX,DY,T).

:- from(X,Y,T), not box(X,Y,T-1).

todo(T) :- target_square(X,Y), step(T), not box(X,Y,T-1).

:- todo(T), not from(X,Y,T) : square(X,Y).

box(X2,Y2,T) :- push(X1,Y1,X2,Y2,DX,DY,T).
box(X1,Y1,T) :- box(X1,Y1,T-1), step(T), not from(X1,Y1,T).

:- target_square(X,Y), steps(T), not box(X,Y,T).

free(X2,Y2,DX,DY,T) :- next(X1,Y1,X2,Y2,DX,DY), from(X1,Y1,DX,DY,T).
free(X2,Y2,DX,DY,T) :- next(X1,Y1,X2,Y2,DX,DY), free(X1,Y1,DX,DY,T),
                       not box(X1,Y1,T).

:- free(X,Y,DX,DY,T), box(X,Y,T-1).

reach(X1,Y1,1) :- initial_at(X1,Y1), step(1).
reach(X1,Y1,T) :- from(X1,Y1,T-1), step(T).
reach(X2,Y2,T) :- next(X1,Y1,X2,Y2,DX,DY), reach(X1,Y1,T), not box(X2,Y2,T-1).

:- next(X1,Y1,X2,Y2,DX,DY), from(X2,Y2,DX,DY,T), not reach(X1,Y1,T).

boxes(N) :- N = #count{X,Y : initial_box(X,Y)}.

:- step(T), boxes(N), #count{X,Y : box(X,Y,T)} != N.
\end{lstlisting}

\paragraph{\textbf{Towers of Hanoi.}}
The baseline encoding for Towers of Hanoi is given in Listing~\ref{lst:encHanoi}.
An instance provides atoms of the form \lstinline{disk/1}, \lstinline{peg/1}, \lstinline{init_on/2}, and \lstinline{goal_on/2}, representing the identifiers of disks or pegs, respectively, as well as the initial and goal pegs for the disks.
Moreover, the atom \lstinline{moves(T)} specifies the maximum number \lstinline{T} of moves. 
The encoding defines the output predicate  \lstinline{move/3}, where \lstinline{move(D,P,T)} means that, at time \lstinline{T}, the disk \lstinline{D} is moved to peg \lstinline{P}. 
The auxiliary predicates \lstinline{time/1}, \lstinline{moved/2}, \lstinline{on/3}, and \lstinline{blocked/3} provide available times, moved disks, the disks' pegs, and disks blocked by some smaller disk on the same peg.
Respective constraints make sure that only the topmost disk of a peg is moved, it is also the smallest disk on its new peg, and the goal pegs are reached at the end of the plan.

\begin{lstlisting}[float=tbph,label=lst:encHanoi,caption={Baseline Encoding for Towers of Hanoi},lastline=17]
time(1..T) :- moves(T).     
on(D,P,0) :- init_on(D,P).                            

{move(D,P,T) : disk(D), peg(P)} = 1 :- time(T).      

moved(D,T) :- move(D,P,T).                                  

blocked(D,P,T) :- on(D,P,T-1), time(T).                    
blocked(D,P,T) :- blocked(D+1,P,T), disk(D).  

:- move(D,P,T), blocked(D,P,T).  
:- moved(D,T), on(D,P,T-1), blocked(D+1,P,T). 

on(D,P,T) :- move(D,P,T).                                      
on(D,P,T) :- on(D,P,T-1), time(T), not moved(D,T).            

:- goal_on(D,P), moves(T), not on(D,P,T).  
\end{lstlisting}

\section{Manual Revision of Streamliners}\label{sec:appendix}

Tables~\ref{tab:manual:pup}, \ref{tab:manual:sokoban}, and \ref{tab:manual:toh}
contrast the streamliners for PUP, Sokoban, or Towers of Hanoi, respectively, with our manual reformulations that are experimentally compared in Section~\ref{subsec:discussion}.
The primary objective of these reformulations is to reduce the grounding size, e.g., by turning constraint P2 / P5 over pairs \lstinline{U1 > U2} of units
into P2r / P5r, where a \lstinline{#count} aggregate lists each unit \lstinline{U} only once (cf.\ Table~\ref{tab:manual:pup}).\footnote{%
The streamliners P2 and P5, generated in separate runs for PUP, are identical and thus have the same reformulation.}
For each sensor~\lstinline{S}, this reformulation yields linear rather than quadratic grounding size in terms of the number of units.
In some cases, our reformulations also strengthen the original streamliners by dropping unnecessary side conditions.
For example, the reformulation S3r of constraint S3 expresses that the target of a push action in Sokoban always contains a box in the next state, while S3 unnecessarily takes another push action into account as well (cf.\ Table~\ref{tab:manual:sokoban}).
Similarly, the reformulations T2r, T4r / T6r, and T5r eliminate redundancies in constraints T2, T4, T5, and T6 for Towers of Hanoi, where the a priori different streamliners T4 and T6 are condensed to the same revised constraint T4r / T6r (cf.\ Table~\ref{tab:manual:toh}).
We do not reformulate constraints P3, P4, and S4 as neither the grounding size nor redundancies can be cut down for them, and T3 is not listed in Table~\ref{tab:manual:toh} because it is a comment that was produced by the LLM.

\begin{table}[b]
\centering
\caption{Manually revised versions of streamliners for PUP\label{tab:manual:pup}}
\begin{tabular}{@{}ll@{}}
\toprule
Constraint & (Re)formulation \\
\midrule
\multirow{2}{*}{P1} &
\lstinline|:-| \lstinline|comUnit(U), #count{Z: unit2zone(U,Z)} = 0,| \\
& \phantom{\lstinline|:-| \lstinline|comUnit(U ),|}%
\lstinline|#count{S: unit2sensor(U,S)} > 0.|
\\\cline{1-2}
P1r &
\lstinline|:-| \lstinline|unit2sensor(U,_), not unit2zone(U,_).| \\\midrule
P2 / P5 &
\lstinline|:-| \lstinline|unit2sensor(U1,S), unit2sensor(U2,S), U1 > U2.| \\\cline{1-2}
P2r / P5r &
\lstinline|:-| \lstinline|sensor(S), #count{U: unit2sensor(U,S)} > 1.| \\\midrule
P3 &
\lstinline|:-| \lstinline|unit2zone(U+1,_), not unit2zone(U,_), comUnit(U), comUnit(U+1).| \\\midrule
\multirow{2}{*}{P4} &
\lstinline|:-| \lstinline|comUnit(U), unit2zone(U,Z1), unit2zone(U,Z2), Z1 < Z2,| \\
& \phantom{\lstinline|:-|} %
\lstinline|#count{S: zone2sensor(Z1,S), zone2sensor(Z2,S)} = 0.| \\\midrule
\multirow{3}{*}{P6} &
\lstinline|:-| \lstinline|comUnit(U1), comUnit(U2), U1 < U2, sensor(S),| \\
& \phantom{\lstinline|:-|} \lstinline|S = #min{S': sensor(S')}, unit2sensor(U2,S),| \\
& \phantom{\lstinline|:-|} \lstinline|#count{Z: unit2zone(U1,Z)} = 0.| \\\cline{1-2}
\multirow{3}{*}{P6r} &
\lstinline|mark(U):-| \lstinline|comUnit(U), unit2sensor(U+1,S), S = #min{X: sensor(X)}.| \\
& \lstinline|mark(U):-| \lstinline|comUnit(U), mark(U+1).| \\
& \lstinline|:-| \lstinline|mark(U), not unit2zone(U,_).| \\
\bottomrule
\end{tabular}
\end{table}

\begin{table}[t]
\centering
\caption{Manually revised versions of streamliners for Sokoban\label{tab:manual:sokoban}}
\begin{tabular}{@{}ll@{}}
\toprule
Constraint & (Re)formulation \\
\midrule
S1 &
\lstinline|:-| \lstinline|push(X1,Y1,X2,Y2,DX,DY,T), push(X3,Y3,X4,Y4,DX,DY,T), X1 < X3.| \\\cline{1-2}
S1r &
\lstinline|:-| \lstinline|diff(DX,DY), step(T), #count{X1 : push(X1,Y1,X2,Y2,DX,DY,T)} > 1.| \\\midrule
\multirow{2}{*}{S2} &
\lstinline|:-| \lstinline|push(X1,Y1,X2,Y2,DX,DY,T), push(X3,Y3,X4,Y4,DX,DY,T),| \\
& 
\phantom{\lstinline|:-| }\lstinline|X1 = X3, Y1 < Y3.| \\\cline{1-2}
\multirow{2}{*}{S2r} &
\lstinline|:-| \lstinline|diff(DX,DY), step(T), square(X,Y),| \\
&
\phantom{\lstinline|:-| }\lstinline|#count{Y1 : push(X,Y1,X2,Y2,DX,DY,T)} > 1.| \\\midrule
\multirow{2}{*}{S3} &
\lstinline|:-| \lstinline|push(X1,Y1,X2,Y2,DX,DY,T), push(X3,Y3,X4,Y4,DX,DY,T+1),| \\
&
\phantom{\lstinline|:-| }\lstinline|X1+Y1 > X3+Y3, not box(X2,Y2,T), X2 != X3, Y2 != Y3.| \\\cline{1-2}
S3r &
\lstinline|:-| \lstinline|push(X1,Y1,X2,Y2,DX,DY,T), not box(X2,Y2,T).| \\\midrule
S4 &
\lstinline|:-| \lstinline|push(X1,Y1,X2,Y2,DX,DY,T), push(X2,Y2,X1,Y1,ODX,ODY,T+1).| \\\midrule
\multirow{2}{*}{S5} &
\lstinline|:-| \lstinline|push(X1,Y1,X2,Y2,DX,DY,T), push(X1,Y1,X3,Y3,DX,DY,T),| \\
& 
\phantom{\lstinline|:-| }\lstinline|X2 = X3, Y2 < Y3.| \\\cline{1-2}
\multirow{2}{*}{S5r} &
\lstinline|:-| \lstinline|path(X1,Y1,X1,Y,DX,DY), step(T),| \\
&
\phantom{\lstinline|:-| }\lstinline|#count{Y2 : push(X1,Y1,X1,Y2,DX,DY,T)} > 1.|
\\\midrule
\multirow{2}{*}{S6} & 
\lstinline|:-| \lstinline|push(X1,Y1,X2,Y2,1,0,T), push(X3,Y3,X4,Y4,-1,0,T),| \\
&
\phantom{\lstinline|:-| }\lstinline|X2 = X4, Y2 < Y4.| \\\cline{1-2}
\multirow{2}{*}{S6r} & 
\lstinline|stop(X2,Y2,DX,0,T):-| \lstinline|push(X1,Y1,X2,Y2,DX,0,T).| \\
&
\lstinline|:-| \lstinline|stop(X,Y1,1,0,T), stop(X,Y2,-1,0,T), Y1 < Y2.| \\
\bottomrule
\end{tabular}
\end{table}

\begin{table}[t]
\centering
\caption{Manually revised versions of streamliners for Towers of Hanoi\label{tab:manual:toh}}
\begin{tabular}{@{}ll@{}}
\toprule
Constraint & (Re)formulation \\
\midrule
T1 &
\lstinline|:-| \lstinline|moved(D1,T-1), moved(D2,T), D2 <= D1, disk(D1+1).| \\\cline{1-2}
\multirow{3}{*}{T1r} &
\lstinline|mark(D,T):-| \lstinline|moved(D,T-1), time(T), disk(D+1).| \\
&
\lstinline|mark(D,T):-| \lstinline|mark(D+1,T), disk(D).| \\
&
\lstinline|:-| \lstinline|mark(D,T), moved(D,T).| \\\midrule
\multirow{4}{*}{T2} &
\lstinline|stable(1,P,T):-| \lstinline|on(1,P,T), goal_on(1,P), time(T).| \\
&
\lstinline|stable(D,P,T):-| \lstinline|on(D,P,T), goal_on(D,P), stable(D-1,P,T),| \\
&
\phantom{\lstinline|stable(D,P,T):-| }\lstinline|disk(D), disk(D-1), time(T).| \\ 
& 
\lstinline|:-| \lstinline|move(D,P,T), stable(D,P_prev,T-1), on(D,P_prev,T-1).| \\\cline{1-2}
\multirow{3}{*}{T2r} &
\lstinline|stable(1,P,T):-| \lstinline|on(1,P,T-1), goal_on(1,P).| \\
&
\lstinline|stable(D,P,T):-| \lstinline|on(D,P,T-1), goal_on(D,P), stable(D-1,P,T).| \\
& 
\lstinline|:-| \lstinline|moved(D,T), stable(D,P,T).| \\\midrule
\multirow{2}{*}{T4} &
\lstinline|:-| \lstinline|disk(D), move(D,P1,T1), move(D,P2,T2), move(D,P1,T3),|
\\
& 
\phantom{\lstinline|:-| }\lstinline|T1 < T2, T2 < T3, T3 = T2 + 1, P1 = P2.| \\\cline{1-2}
T4r / T6r &
\lstinline|:-| \lstinline|move(D,P,T), move(D,P,T+1).| \\\midrule
\multirow{2}{*}{T5} &
\lstinline|:-| \lstinline|move(D1,P1,T), move(D2,P2,T+1), D1 < D2,| \\
&
\phantom{\lstinline|:-| }\lstinline|on(D1,Px,T-1), on(D2,Py,T), Px = Py, P1 != Py, P2 = Px.| \\\cline{1-2}
T5r &
\lstinline|:-| \lstinline|move(D,P,T), on(D,P,T-1).| \\\midrule
\multirow{2}{*}{T6} &
\lstinline|:-| \lstinline|time(T), T > 2, move(D,P1,T-1), move(D,P2,T-2), P1 = P2,| \\
& 
\phantom{\lstinline|:-| }\lstinline|move(D,P1,T).| \\
\bottomrule
\end{tabular}
\end{table}

\end{document}